# TriMat: Context-aware Recommendation by Tri-Matrix Factorization

Hao Wang
aRatidar.com, Beijing, China
* Corresponding author: haow85@live.com

## ABSTRACT

Search engine is the symbolic technology of Web 2.0, and many people used to believe recommender systems is the new frontier of Web 3.0. In the past 10 years, with the advent of TikTok and similar apps, recommender systems has materialized the vision of the machine learning pioneers. However, many research topics of the field remain unfixed until today. One such topic is CARS (Context-aware Recommender Systems) , which is largely a theoretical topic without much advance in real-world applications. In this paper, we utilize tri-matrix factorization technique to incorporate contextual information into our matrix factorization framework, and prove that our technique is effective in improving both the accuracy and fairness metrics in our experiments.

## 1. INTRODUCTION

As the symbolic technology of Web 3.0, recommender system has emerged as one of the most successful technologies in the modern day internet age. By recruiting a small team of a dozen engineers, many companies have been able to produce cutting edge recommender system technologies and products that yield high ROI for the organization. There are many practical scenarios in which recommendation technologies could be applied to. Famous companies such as TikTok, Amazon, Kuai Shou and Netflix are just a few examples of organizations that succeed in the technology commercialization.

The major goal of recommender system technology is to increase the accuracy of the algorithmic output. We want to predict the user preference of unseen items precisely so more people will visit the website or purchase our products. The initial metric used to design the algorithm is RMSE and MAE. Later, other approaches such as learning to rank have been invented, and the algorithmic evaluation metrics have been diversified to include AUC, NDCG and other information retrieval measures. Since the year of 2016, a new technological trend marked by deep learning [1][2] has become the new de facto standard of recommendation algorithm design.

Although research of recommender system has been going on for decades, there have been several practical problems associated with the field that remain unanswered. One such problem is CARS (Context-aware Recommender Systems). In addition to the user item rating data associated with recommender systems, there are contextual information fields that are also extremely helpful. CARS is the technology that takes advantage of not only the user item rating data, but also contextual information field. Although CARS is widely applicable in scenarios such as in-car music recommendation, travel guide recommendation, fast-food recommendation, etc.

In 2021, H. Wang proposed a CARS algorithm named MatMat [3], which reduces the CARS problem into a matrix factorization by matrix fitting algorithm. In his following publication, H. Wang introduced a practical example of MatMat [3] named MovieMat [4] that creates a movie recommendation algorithm.

In this paper, we propose a new algorithm that uses tri-matrix factorization to solve the CARS problem. We prove in experiments that our algorithm is competitive with MatMat / MovieMat framework / technique. We hope our work could help improve the current status quo of CARS research and find solutions to make theories become practice.

## 2. RELATED WORK

Recommender systems is a field that boasts rich history and deep scholarship. Early models such as user-based collaborative filtering [5] and item-based collaborative filtering [6] are widely applied in the industry. More complex models such as matrix factorization have witnessed an explosion of inventions in a brief period of time: SVD++ [7], SVDFeature [8], timeSVD [9], ALS [10] are just several few examples of the broad spectrum of matrix factorization techniques. In just a very short time, another milestone algorithm named Factorization Machines [11] have been invented with variants such as FFM [12] following suit. The techniques have been adopted widely in companies such as Meituan.com.

Learning to rank was the next milestone in the history of recommender systems. Since the early models of BPR [13] and CLiMF [14], later models have become more versatile in solving a wider range of problems especially fairness [15][16][17]. In the same time, linear models and hybrid models are also popular in the industry. Companies such as Baidu.com [18][19] and Netease.com [20] have put into practice large scale models to tackle real world business problems.

Since 2016, a new technological trend emerged as deep learning infiltrated into the machine learning arena. Algorithmic approaches such as DeepFM [21], Wide & Deep [22], DCN [23], AutoRec [24], AutoInt [25], among a whole large repository of new inventions. Deep learning approaches are highly effective in industrial contexts. YouTube [26] and other corporations have invested a large amount of resources on the topic.

Another topic that has been heavily researched in recent years is fairness problem. H. Wang invented an algorithm named MatRec [27] that models the fairness variable into the matrix factorization paradigm. In the following year, Zipf Matrix Factorization [28] and KL-Mat [29] were also proposed by the same author, utilizing regularization technique to solve the fairness problem. E. Chi [30] in 2017 proposed a fairness algorithm named Focused Learning. More research on the topic emerged in top conferences such as SIGIR [31][32] since then, using learning to rank as the major research benchmark.

Cold-start problem is yet another research topic that arouses research interest among researchers and industrial workers. In order to solve the problem, researchers have taken advantage of techniques such as meta learning [33] and transfer learning [34] to alleviate the issue. In 2021, H. Wang proposed a new set of algorithms that used Zeroshot Learning to mitigate the problem. The algorithms include ZeroMat / ZeroMat Hybrid [35], DotMat / DotMat Hybrid [36], and PowerMat [37]. These algorithms rely on no user item rating data for the recommendation task.

## 3. TRI-FACTORIZATION BASED CARS

We create our new CARS algorithm based on matrix factorization. The formal definition of matrix factorization is defined in the following way :

$$L = \sum_{i=1}^{n} \sum_{i=1}^{m} \left( R_{i,j} - U_i^T \cdot V_j \right)^2$$

Normally, the loss function needs to be redefined to avoid numerical problems related to the computation :

$$L = \sum_{i=1}^{n} \sum_{i=1}^{m} \left( \frac{R_{i,j}}{R_{max}} - \frac{U_i^T \cdot V_j}{||U_i|| \times ||V_j||} \right)^2$$

The basic idea behind matrix factorization is to approximate the user item rating values with dot products of vectors in higher dimensions. In this way, both the time and space consumption of matrix factorization algorithm can be minimized to a much smaller scale than the $O(n^2)$ or $O(n^3)$ scale related to matrix factorization algorithm in the numerical analysis sense.

Instead of factorizing the user item rating matrix into multiplication of 2 matrices, we propose to factorize the user item rating matrix into 3 matrices, and we name the algorithm Tri-factorization based CARS (TriMat). Formally, we define the algorithm loss function as below :

$$L = (R - U^T \cdot C \cdot V)^2$$

In this formulation, U and V are defined in the sense as in the classic matrix factorization, and c is the context matrix that contains the contextual information. To solve for the optimal parameters , we resort to Stochastic Gradient Descent (SGD) and obtain the following update rules for parameters U and V :

$$\frac{\partial L}{\partial U} = -2(R - U^T \cdot C \cdot V)C \cdot V$$
$$\frac{\partial L}{\partial V} = -2(R - U^T \cdot C \cdot V)C^T \cdot U$$
$$\frac{\partial L}{\partial C} = -2(R - U^T \cdot C \cdot V)U \cdot V^T$$

Please notice the counter-intuitive definition of C : C is only a constant matrix at the beginning, and afterwards its value is updated together with U and V.

In the experiment section, we use LDOS-CoMoDa dataset to test our algorithm. We set the contextual information matrix C as follows, and define the shape of U and V accordingly :

$$C = \begin{bmatrix} \frac{location}{max(location)} & \frac{mood}{max(mood)} \\ \frac{weather}{max(weather)} & \frac{season}{max(season)} \\ \frac{datetype}{max(datetype)} & \frac{emo}{max(emo)} \end{bmatrix}$$

Please notice the analogy between TriMat and SVD decomposition. Both algorithms decompose the user item rating matrix into 3 matrices. The difference is the contextual information matrix is dense while the singular matrix is diagonal. We believe a dense contextual information matrix could represent more knowledge than the diagonal structure. The side effect of the algorithm is the limit imposed to the dimension size of the U and V vectors. Since contextual information matrix size is usually on a much smaller scale than the number of users or the number of items, it does have limitation on the shapes of U and V since they need to comply with the shape of C. However, a nice property introduced by the algorithm is the reduction on the space consumption of the algorithm. The space complexity of TriMat is only a fraction of the space complexity of the classic matrix factorization and other CARS algorithms (*Context-aware Recommender System* algorithms).

To be more precise on the advantages of TriMat algorithm : The dimensions of U and V in the algorithmic example in this section are 3 and 2 respectively, while the classic matrix factorization might have a dimension of 30 or even higher. TriMat usually consumes less than 10% of the total space consumption of other matrix factorization paradigms. We demonstrate that by only using less than 10% of the space consumed by other algorithms, TriMat is able to produce competitive results. The implication of the algorithm and results is that TriMat is a competitive small AI model that is not only suitable for computers but also other devices that run embedded systems.

## 4. CARS AND TINYML

One important application scenario of CARS is embedded systems. By incorporating CARS into automobiles' music audio systems, or theaters' sensor systems (as discussed in the MovieMat paper [4]), we are able to improve the user experiences of conventional industry products and services. But there is an intrinsic problem related to embedded systems : The demand for space and time complexity is very restrictive and many programs that run on computers can not be imported to run on the embedded systems.

However, embedded systems is so important for CARS applications that we could not ignore this field. TinyML has emerged in recent years to reduce the complexity of large AI models to only a fraction of its original size. Our proposed model is an example of small AI model that does not require TinyML techniques such as pruning and quantization to reduce the model size.

One interesting research topic is the optimal size of the dimensions of U and V. So far there has been no rigorous research results on the problem. We are also astonished by the effectiveness of small dimension sizes in TriMat algorithm. One of our future research directions is the reason behind this phenomenon.

## 5. EXPERIMENT

LDOS-CoMoDa Dataset [38] comprises of 121 users and 1232 movies. The contextual information matrix used in TriMat in our experiment is defined as in the previous section. We compare TriMat against the classic matrix factorization and MovieMat :

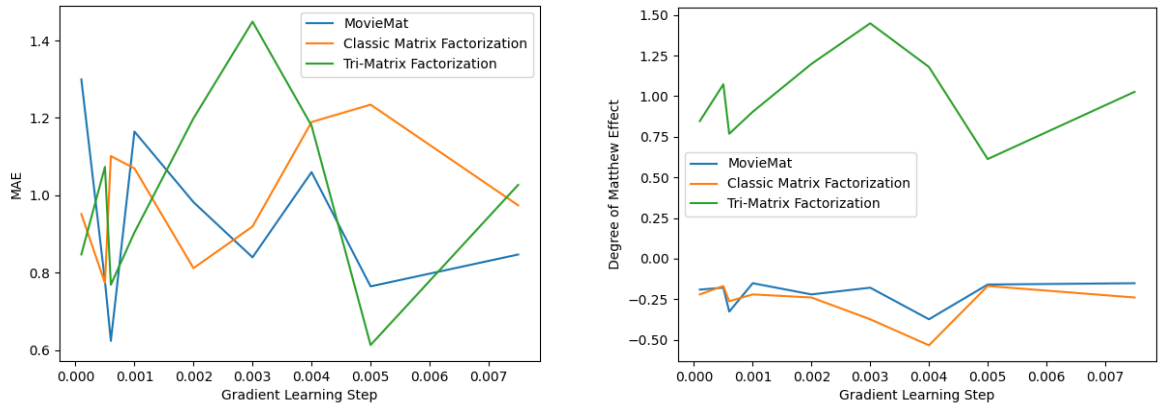

Fig.1 LDOS-CoMoDa Dataset Experiment

From the experimental results shown in Fig. 1, we observe that the best MAE achieved by TriMat is competitive with MovieMat when we do Grid Search on the step learning rate. We also observe that TriMat produces much fairer results than MovieMat when compared on Degree of Matthew Effect [28]. Degree of Matthew Effect, as defined in [28] defines the degree of popularity bias in a statistical measure that is widely adopted by the same author in his publications.

Since it is extremely difficult to acquire open datasets for CARS online, we have not been able to find another dataset to test our algorithms. So we convince ourselves that by testing on LDOS-CoMoDa with MovieMat and classic matrix factorization, we have been able to prove the competitiveness of our algorithm.

# 6. CONCLUSION

In this paper, we proposed a new CARS (context-aware recommender system) algorithm named TriMat. The algorithm decomposes the user item rating matrix into 3 matrices and uses Stochastic Gradient Descent to optimize the parameters. Our algorithm is competitive with the cutting edge CARS algorithm MovieMat on both the accuracy and fairness metrics. TriMat is a small AI model that does not require TinyML techniques such as pruning or quantization to reduce the complexity of the model size.

In future work, we would like to explore the interaction among U, V and C - the 3 matrices after decomposition. We would like to find other CARS datasets for testing as well, so we can further validate the legitimacy of our algorithm. We are also interested in exploring the theory behind the validity of TriMat : Why there exists small models that are competitive with large models ? In addition, the Degree of Matthew Effect is usually negative for most algorithms. However, TriMat has a positive Degree of Matthew Effect. What is the implication of the phenomenon ?

# ACKNOWLEDGEMENT

I am grateful to the internet age - Tens of millions of computer scientists and engineers have created the greatest feat of our time - the internet industry. Computing facilities such as QingCloud.com and Lenovo laptop have made everything possible for my independent research. Axiv.org and other publication repositories have made knowledge flow so smooth and convenient, that almost every paper that I want to read can be accessed on the internet.

I landed 3 research awards at international conferences by 2021. In 2021, I also picked up my course in English competition and entered the National Final on the professional track of a national English contest with 81,000 contestants. I am achieving recognition that I was able to achieve around 18 years ago. I am not content with my current academic and professional excellence. I want to earn much more in both monetary and academic terms. In spite of this, I am grateful to my own persistence and hardwork, without which I won't be able to be who I am today.